\documentclass[prl,aps,twocolumn,showpacs,superscriptaddress]{revtex4-1}
\usepackage{amssymb}
\usepackage{amsmath}
\usepackage{graphicx}
\usepackage{dcolumn}
\usepackage{bm}
\usepackage{float}
\usepackage[colorlinks,
            linkcolor=blue,
            anchorcolor=blue,
            citecolor=blue
            ]{hyperref}
\begin{document}


\title{Prediction of Novel Stable 2D-Silicon with Fivefold Coordination}

\author{Zhenhai Wang}
\email{wangzh@njupt.edu.cn (Z.W.)}
\affiliation{Nanjing University of Posts and Telecommunications, Nanjing 210003, China}
\affiliation{Department of Geosciences, Center for Materials by Design, and Institute for Advanced Computational Science, Stony Brook University, Stony Brook, NY 11794, USA}


\author{Mingwen Zhao}
\email{zmw@sdu.edu.cn (M.Z.)}
\affiliation{School of Physics and State Key Laboratory of Crystal Materials, Shandong University, Jinan 250100, China}

\author{Xiang-Feng Zhou}
\affiliation{Department of Geosciences, Center for Materials by Design, and Institute for Advanced Computational Science, Stony Brook University, Stony Brook, NY 11794, USA}
\affiliation{School of physics and Key Laboratory of Weak-Light Nonlinear Photonics, Nankai University, Tianjin 300071, China}

\author{Qiang Zhu}
\affiliation{Department of Geosciences, Center for Materials by Design, and Institute for Advanced Computational Science, Stony Brook University, Stony Brook, NY 11794, USA}

\author{Xiaoming Zhang}
\affiliation{School of Physics and State Key Laboratory of Crystal Materials, Shandong University, Jinan 250100, China}

\author{Huafeng Dong}
\affiliation{Department of Geosciences, Center for Materials by Design, and Institute for Advanced Computational Science, Stony Brook University, Stony Brook, NY 11794, USA}

\author{Artem R. Oganov}
\email{artem.oganov@stonybrook.edu (A.R.O.)}
\affiliation{Department of Geosciences, Center for Materials by Design, and Institute for Advanced Computational Science, Stony Brook University, Stony Brook, NY 11794, USA}
\affiliation{Skolkovo Institute of Science and Technology, Skolkovo Innovation Center, 3 Nobel Street, Moscow 143026, Russia}
\affiliation{Moscow Institute of Physics and Technology, 9 Institutskiy Lane, Dolgoprudny City, Moscow Region 141700, Russia}
\affiliation{School of Materials Science, Northwestern Polytechnical University, Xi'an 710072, China}

\author{Shumin He}
\affiliation{Nanjing University of Posts and Telecommunications, Nanjing 210003, China}

\author{Peter Gr\"{u}nberg}
\affiliation{Peter Gr\"{u}nberg Institute, Forschungszentrum J\"{u}lich, Wilhelm-Johnen-Stra$\beta$e, J\"{u}lich 52428, Germany }

\begin{abstract}
Silicene, an analogue of graphene, was so far predicted to be the only two-dimensional silicon (2D-Si) with massless Dirac fermions. Here we predict a brand new 2D-Si Dirac semimetal, which we name siliconeet [\textit{silik'ni:t}]. Unexpectedly, it has a much lower energy than silicene and robust direction-dependent Dirac cones with Fermi velocities comparable to those in graphene. Remarkably, its peculiar structure based on pentagonal rings and fivefold coordination plays a critical role in the novel electronic properties. Taking spin-orbit coupling into account, siliconeet can also be recognized as a 2D-topological insulator with a larger nontrivial band gap than silicene.
\end{abstract}

\maketitle


Since experimental realization in 2004 \cite{R01}, the most stable two dimensional (2D) form of carbon, graphene, has risen as the hottest 2D material \cite{R02}. Its unique band structure gives graphene massless Dirac fermions, leading to many novel physics, such as high carrier mobility \cite{R03} and quantum Hall effect \cite{R04,R05,R06}. Silicon (Si), as the group-IV element adjacent to carbon in the Periodic Table, should also be explored in the 2D-state: realization of 2D forms will make Si, the leading material of microelectronic technology, much more promising for future electronic device applications.

Unlike carbon, bulk Si cannot form a natural layered phase like graphite. Thus, it is impossible to obtain 2D-Si by simple mechanical exfoliation \cite{R01}. Silicene, with a graphene-like honeycomb structure, was so far predicted to be the only 2D-Si structure with Dirac fermions \cite{R07,R08,R09}. Its novel properties related to Dirac linear dispersion, such as quantum spin Hall (QSH) effect \cite{R10}, chiral superconductivity \cite{R11}, giant magnetoresistance \cite{R12}, and various exotic filed-dependent states \cite{R13}, \emph{etc.}, have been proposed and theoretically explored. Distinct from graphene with \emph{sp}$^{2}$-bonding planar structure, silicene favors a low-buckling configuration with a mixed \emph{sp}$^{2}$-\emph{sp}$^{3}$ like hybridized state \cite{R07,R09,R14}. Experimentally, silicene was usually deposited on substrates such as Ag(111) \cite{R15,R16,R17}, Ag(110) \cite{R18,R19}, Ir(111) \cite{R20}, ZrB$_2$(0001) \cite{R21}, and MoS$_2$ \cite{R22}. Mixed \emph{sp}$^{2}$-\emph{sp}$^{3}$ hybridized state also makes silicene strongly interact with substrates \cite{R23,R24,R25}, leading to ambiguous electronic properties in practical observations. Moreover, silicene proved unstable in the air due to sensitive surface states \cite{R26}, which are also derived from the mixed \emph{sp}$^{2}$-\emph{sp}$^{3}$ orbital. All this makes it difficult to transfer honeycomb silicene into free-standing state. Up to now, no direct evidence shows Dirac states in 2D-Si.

In contrast to honeycomb graphene and silicene, predictions of 6, 6, 12-graphyne \cite{R27}, \emph{Pmmn}-boron \cite{R28} and phagraphene \cite{R29} all have suggested that hexagonal symmetry is not a prerequisite for the appearance of Dirac cones. In this Letter, using evolutionary structure searching, we predict a brand new 2D-Si Dirac allotrope, which we name siliconeet. Compared with honeycomb silicene, this peculiar structure has a much lower energy and robust direction-dependent Dirac cones. Further investigations confirm that the pentagonal atomic rings with fivefold coordination play a critical role in its novel electronic properties. Taking spin-orbit coupling (SOC) into account, siliconeet is also confirmed to be a 2D topological insulator with a larger SOC gap than silicene \cite{R10}.

We performed systematic 2D structure searches via the \emph{ab initio} evolutionary algorithm USPEX \cite{R30,R31,R32} considering 6, 8, 10, 12, 14, 16, 18, 20, 22 and 24 Si atoms per unit cell. The produced structures were all relaxed, and their energies were used for selecting structures as parents for the new generation of structures. Structure relaxations and total energy calculations were performed using projector-augmented wave (PAW) method \cite{R33}, as implemented in the Vienna \emph{ab initio} simulation package (VASP) \cite{R34,R35}. Exchange-correlation energy was treated within the generalized gradient approximation (GGA), using the functional of Perdew, Burke, and Ernzerhof (PBE) \cite{R36} and more accurate hybrid functional HSE06 \cite{R37,R38}. Brillouin Zone (BZ) integrations were carried out using $\Gamma-$centered sampling grids with resolution of 2$\pi$$\times$0.02 {\AA}$^{-1}$ for final structure optimizations. Atomic positions and lattice constants were optimized using the conjugate gradients (CG) scheme. A kinetic energy cutoff of 600 eV was adopted and the threshold of electronic self-consistency was set as 10$^{-8}$ eV. Dense k-point grid (30$\times$30$\times$1) was employed for charge density and band structure calculations. Phonon calculations by supercell approach using the Phonopy code \cite{R39} and first-principles molecular dynamics simulations under constant temperature and volume (NVT) were performed to examine dynamical stability with respect to infinitesimal and finite distortions.

\begin{figure}
\begin{center}
\includegraphics[width=1.0\columnwidth]{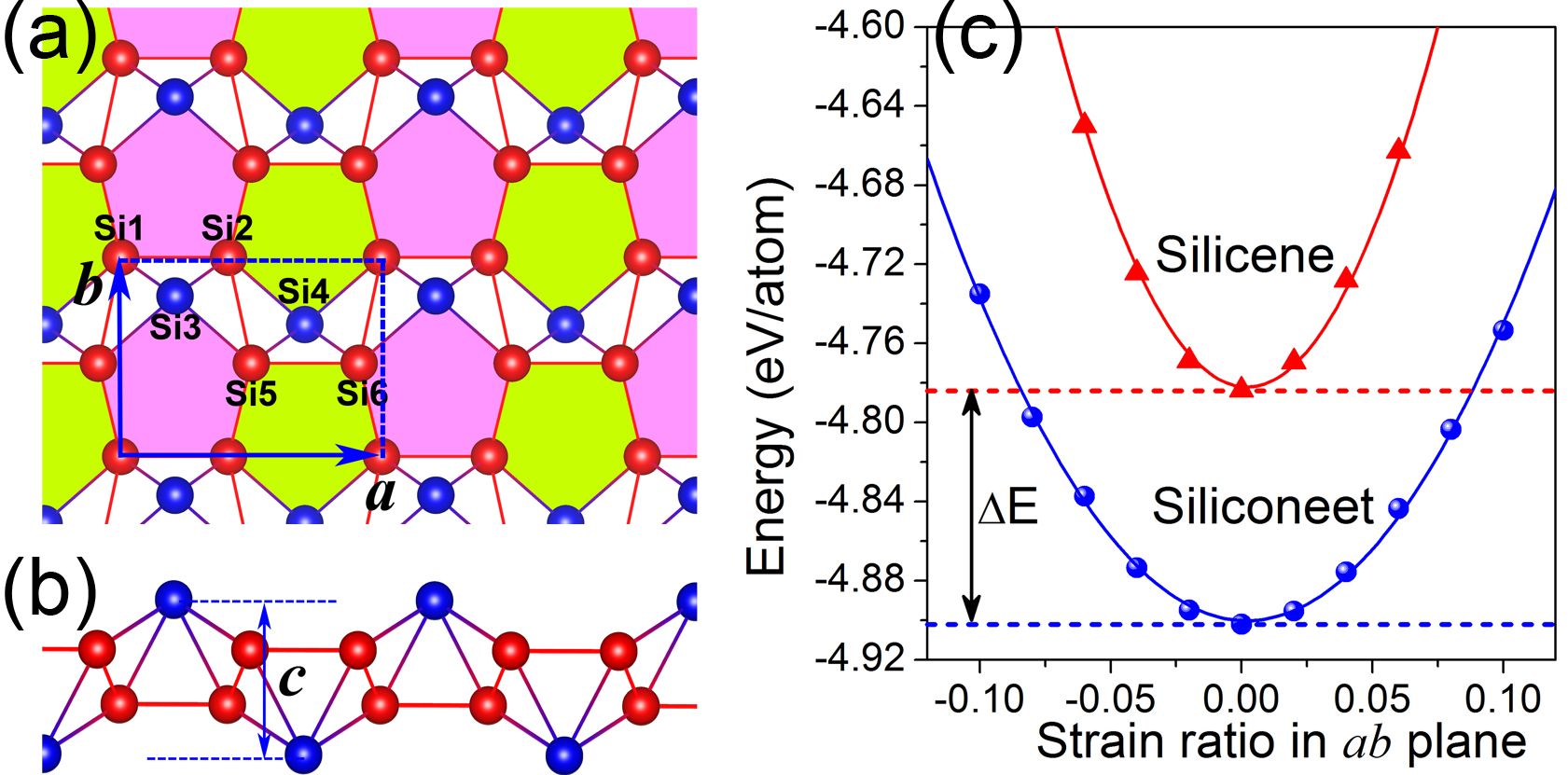}
\end{center}
\caption{\label{fig1}
(color online) [(a) and (b)] Top and side geometric views of the 2D-Si nanosheet (we name siliconeet). Top- and bottom-Si atoms are blue, while middle-Si atoms are red. Si1-Si6 are atoms in its unit cell. (c) Energy of silicene and siliconeet under strain in \emph{ab} plane,  $\triangle$E = 0.119 eV/atom based on GGA-PBE results. }
\end{figure}

In this work, many 2D-Si structures were generated. Most structures with low energies have disordered geometries. Interestingly, pentagonal rings could frequently be found among them. Siliconeet, a peculiar 2D-structure containing pentagonal rings, is shown in Fig. \ref{fig1}(a) and \ref{fig1}(b). Its rectangular lattice constants are \emph{a} = 5.544 {\AA} and \emph{b} = 4.238 {\AA}, with a thickness \emph{c} = 3.290 {\AA}. In fact, silicene is also non-planar with a 0.450 {\AA} buckling as discussed above. Silicene's honeycomb structure has only one independent atomic site and the Si-Si bond length is 2.278 {\AA}. Siliconeet is quite different [Fig. \ref{fig1}(a)], with 2 nonequivalent atoms (Si1 and Si3) and 4 distinct Si-Si bonds with bond lengths of 2.294 {\AA} (Si1-Si2), 2.412 {\AA} (Si2-Si4, Si3-Si5), 2.582 {\AA} (Si2-Si5) and 2.637 {\AA} (Si1-Si3, Si2-Si3), respectively. Siliconeet structure can be considered as an elongated honeycomb silicene with adatoms (top-Si3 and bottom-Si4), as shown in Fig. S1. Its side-view looks like recently reported 2D Dirac \emph{Pmmn}-boron \cite{R28}, and it also has two sub-lattices (blue top-, bottom-Si atoms and red middle-Si atoms), as shown in Fig. \ref{fig1}(b).

Prior work confirmed that buckled 2D boron structures are usually more stable than planar ones, becuase \emph{sp}$^{2}$-hybridization is hard to achieve, and the same is true for 2D-Si. Total energy calculations show that siliconeet has lower energy than silicene even under external strain up to 8.8\%, as shown in Fig. \ref{fig1}(c). The energy difference is as large as 119 meV/atom based on GGA-PBE calculations. Indeed, we see how far siliconeet is from \emph{sp}$^2$-hybridization: (1) The hypothesis adsorbing atoms on elongated honeycomb silicene, allow a good release of the instability of \emph{sp}$^2$-like bonding configurations; (2) Si pentagonal rings are formed when adsorbing atoms on the middle sublattice. Strikingly, this peculiar structure with pentagonal rings presents two different coordination numbers, fivefold for the middle and fourfold for the adsorbing surface atoms.

\begin{figure}
\begin{center}
\includegraphics[width=1.0\columnwidth]{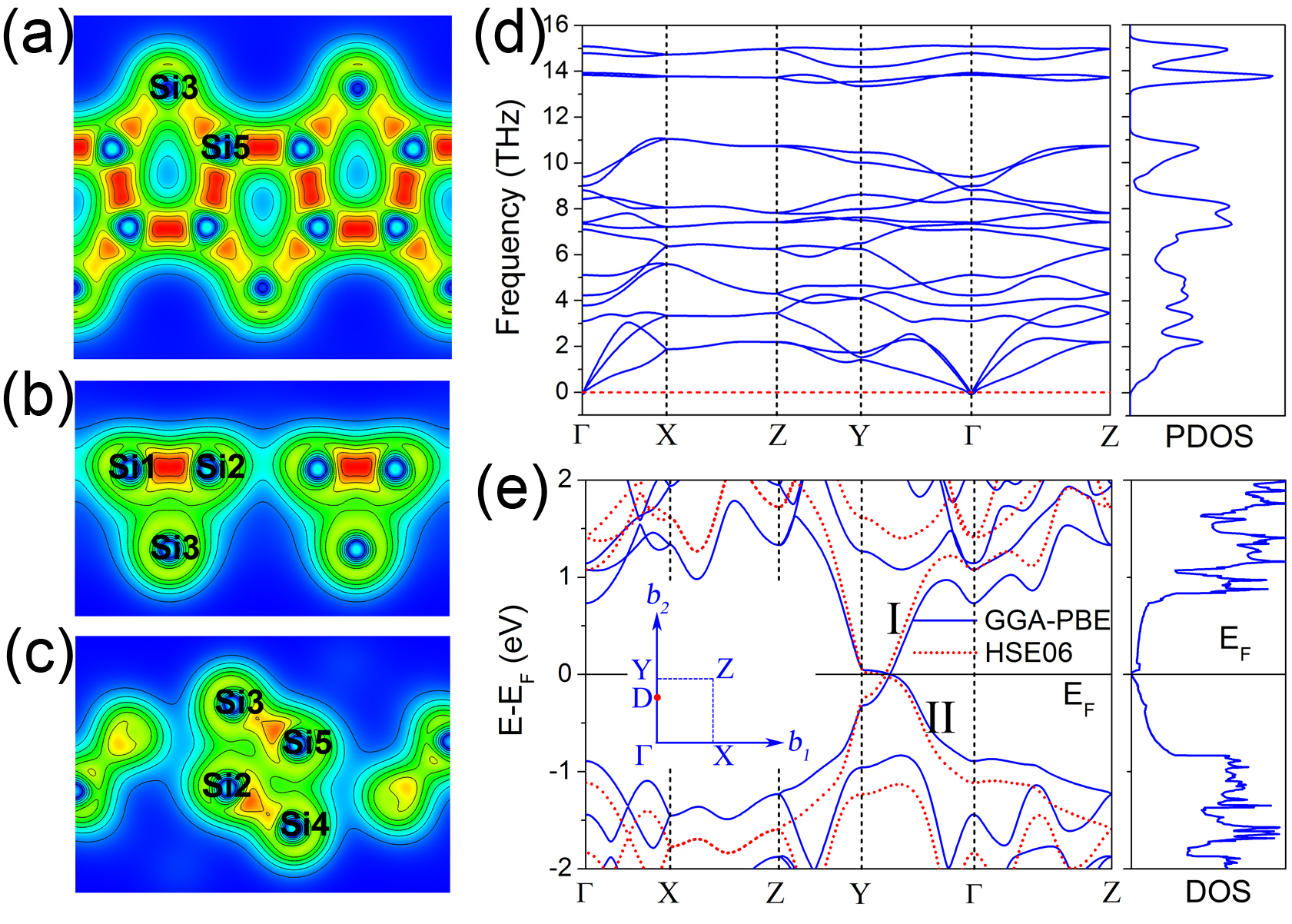}
\end{center}
\caption{\label{fig2}
(color online) Charge density distributions in different sections: (a) pentagonal ring section, (b) Si1-Si2-Si3 section, (c) Si2-Si3-Si5 section. (d) Phonon dispersion and phonon density of states (PDOS). (e) Electronic band structures and density of states (DOS). Band structures are calculated using the GGA-PBE and HSE06 functionals, respectively. The Fermi level is set to zero.}
\end{figure}

Charge density distributions in Fig. \ref{fig2}(a)-(c), which are projected to different cross-sections around top-Si atoms, further validate its bonding configurations. Obviously, top-Si (Si3 as example) and atoms from the middle sublattice (Si5 as example) have two and three strong covalent bonds in pentagonal rings respectively, as shown in Fig. \ref{fig2}(a). Meanwhile, as depicted in Fig. \ref{fig2}(b) and \ref{fig2}(c), Si3 has obvious charge density overlaps with atoms (Si1 and Si2) in the neighboring pentagonal ring, while Si5 has overlaps with Si2 and Si4. In this way, top- and bottom-Si (Si3 and Si4) present fourfold coordination and middle ones (Si1, Si2, Si5 and Si6) fivefold. Note that stable bulk Si has fourfold coordination \cite{R40,R41}, while threefold in silicene. Here is the first time to demonstrate fivefold coordination can be formed in periodic 2D-Si structure at normal conditions.

For this peculiar 2D-Si structure with fivefold coordination, we confirmed the dynamical stability by checking its phonon dispersion curves and phonon density of states (PDOS), which show no imaginary frequencies, as presented in Fig. \ref{fig2}(d). To examine its thermal stability, a 3$\times$3 supercell was built to perform \emph{ab initio} molecular dynamics simulations. After heating at room temperature (300 K) for 3 \emph{ps} with a time step of 1 \emph{fs}, no structural change occurred. We also have verified it using a fixed-cell technique in USPEX with lattice-matching Ag(110) substrate, as shown in Fig. S2. Siliconeet is located at the lowest energy region of the structure evolution, which implies the possibility of growing it on Ag(110) substrate. All in all, siliconeet can exist with peculiar pentagonal rings, and is much more stable than honeycomb silicene due to the presence of both fivefold and fourfold coordinations.

 Siliconeet is not only lower in energy than silicene, but also has unique electronic properties. The band structures in Fig. \ref{fig2}(e) show a distorted Dirac cone in the rectangular first BZ. The valence and conduction bands (denoted as bands I and II) meet at the Fermi level and form distorted Dirac cones. The density of states (DOS) is zero at the Fermi level, i.e. siliconeet is a semimetal, which further supports the presence of Dirac cones. Fermi velocities ($\upsilon_\emph{f}$) in both \emph{k}$_\emph{x}$ and \emph{k}$_\emph{y}$ directions were obtained from slopes of the bands at the Dirac point. In the \emph{k}$_\emph{x}$ direction, $\partial$\emph{E}/$\partial$\emph{k}$_\emph{x}$ = $\pm$34.29 eV{\AA} ($\upsilon$$_\emph{fx}$ = 8.29$\times$10$^5$m/s); while in the \emph{k}$_\emph{y}$ direction, the slope of the bands equals -3.74 eV{\AA} ( $\upsilon$$_\emph{fy}$ = -0.91$\times$10$^5$m/s) and 18.05 eV{\AA} ($\upsilon$$_\emph{fy}$ = 4.36$\times$10$^5$m/s). Note that $\upsilon$$_\emph{fx}$ of this direction-dependent Dirac cone is even larger than that of graphene ($\upsilon$$_\emph{f}$ = 8.22¡Á$\times$10$^5$m/s) \cite{R29} based on GGA-PBE results.

\begin{figure}
\begin{center}
\includegraphics[width=1.0\columnwidth]{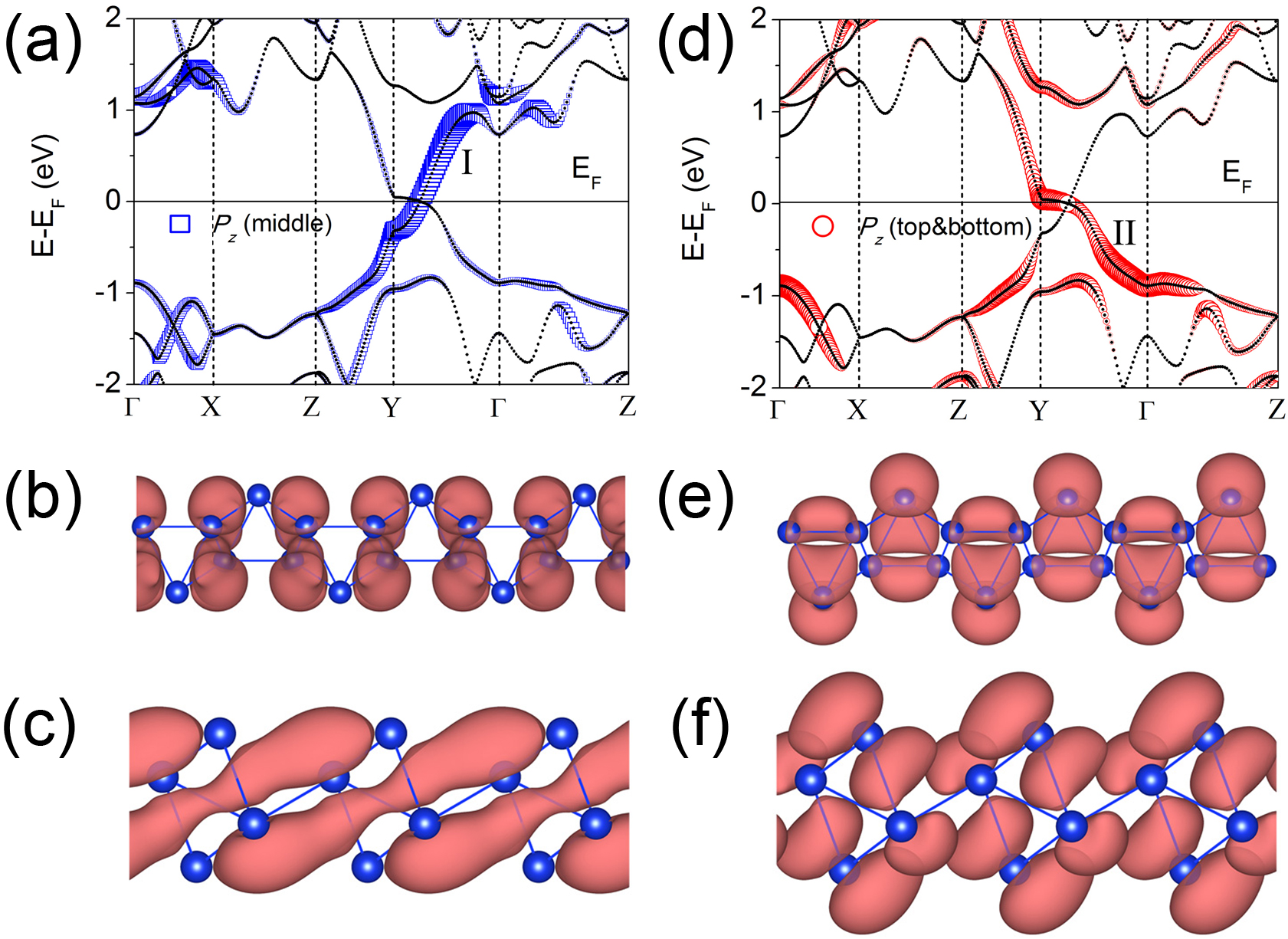}
\end{center}
\caption{\label{fig3}
(color online) $\emph{P}_z$-orbital resolved band structures and partial charge densities for [(a)-(c)] band-I and [(d)-(f)] band-II.  [(b) and (e)] are views of \emph{ac} plane, [(c) and (f)] are views of \emph{bc} plane. In (a) and (d), Fermi level is set to zero.}
\end{figure}

\begin{figure}
\begin{center}
\includegraphics[width=0.9\columnwidth]{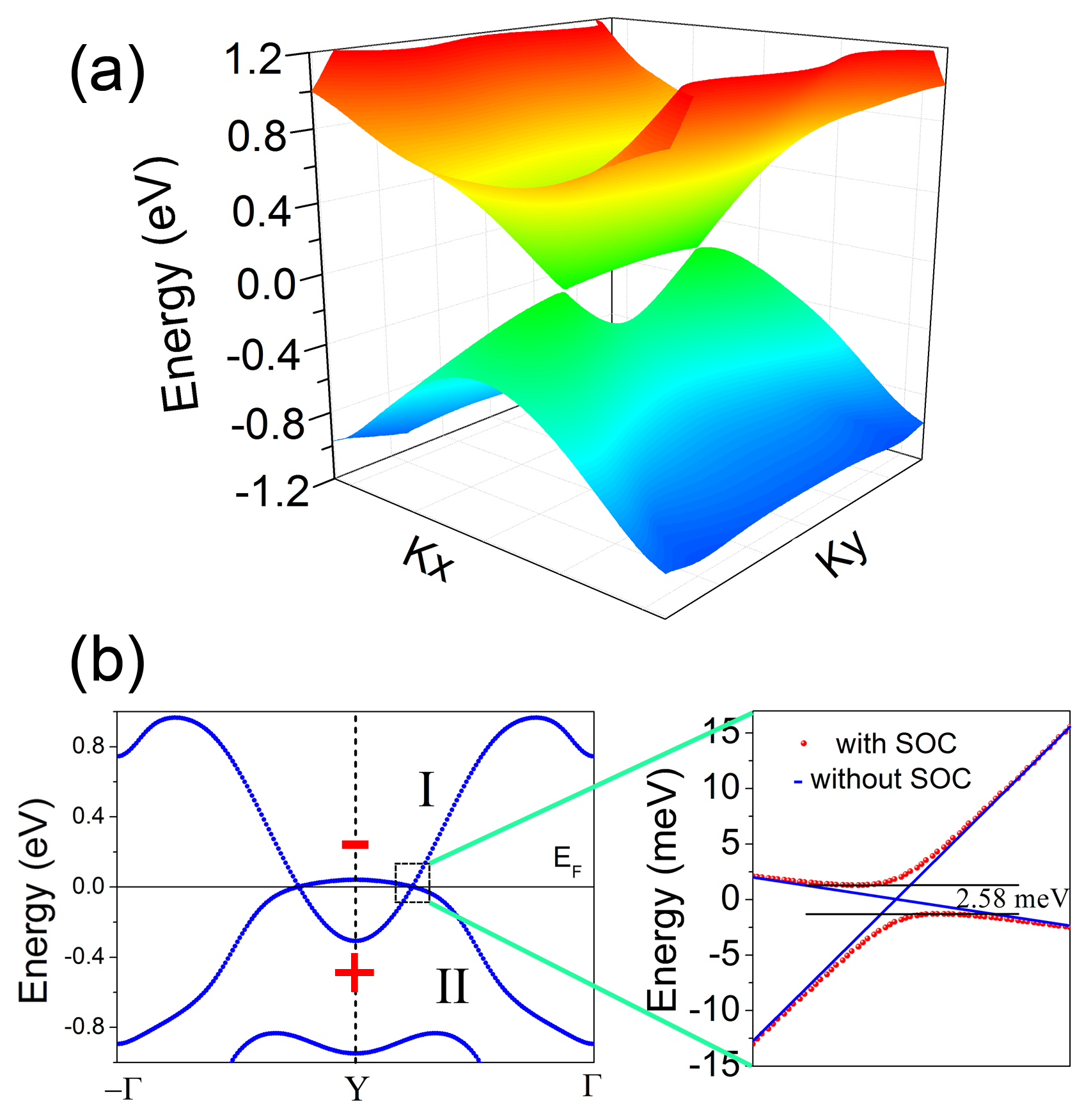}
\end{center}
\caption{\label{fig4}
(color online) (a) Distorted Dirac cones formed by valence and conduction bands in the vicinity of the Dirac points. (b) Band structure along the -$\Gamma-Y-\Gamma$ line, parities (even, odd) of the two bands near the Fermi level at the Y point have been denoted by (+, -). At right is band detail in the vicinity of the distorted Dirac cone with and without spin-orbit coupling (SOC), the Fermi level is set to zero.}
\end{figure}

To further explore the origin of Dirac bands, orbital-resolved band structures and partial charge density calculations for the two bands near the Fermi level were performed, which are depicted in Fig. \ref{fig3}. It is obvious that they are mainly from the \emph{p}$_\emph{z}$ orbitals of the two sublattices. Interestingly, due to tilting of pentagonal rings (about 36$^\circ$ with respect to the \emph{ab} plane), \emph{p}$_\emph{z}$ orbitals also have a tilting angle, as shown in Fig. \ref{fig3}(c) and \ref{fig3}(f). While band-I is mainly composed of \emph{p}$_\emph{z}$ orbitals of the middle sublattice, band-II mainly originates from \emph{p}$_\emph{z}$ orbitals of surface atoms. Moreover, the dispersion of band-I is much larger than that of band-II near Fermi level, which results in direction-dependent Fermi velocities. Such unique Dirac bands are presented in a more pictorial way in Fig. \ref{fig4}(a), and the corresponding bands along the -$\Gamma-Y-\Gamma$ line are depicted in Fig. \ref{fig4}(b). The origin of the direction-dependent Dirac cones, as we have discussed in detail in our recent ``phagraphene" work \cite{R29}, is from the band crossing in the vicinity of the Fermi level: only bands I and II can appear and invert to each other at the Fermi level, Dirac cones are then produced on both sides of the space-symmetric and time-invariant \emph{k}-point Y (or $\Gamma$). These features proved to be robust and can be kept under 6\% tensile and compressive strain, as shown in Fig. S3.

Another interesting property of graphene and silicene is their topologically nontrivial electronic band structures. Taking spin-orbit coupling (SOC) into account, graphene was proposed to be a 2D topological insulator with a nontrivial bulk band gap and gapless edge states protected by time-reversal symmetry and responsible for the QSH effect \cite{R04}. However, the weak SOC of carbon and planar configuration of graphene lead to an unobservable small bulk band gap ($\thicksim$$10^{-3}$ meV) at the Dirac points. Thanks to the greater intrinsic SOC strength of silicon atoms and the buckled configuration, the SOC gap in silicene is greatly increased to 1.55 meV \cite{R10}. Here we confirmed the topological nontriviality of siliconeet from the nonzero topological invariant $\emph{Z}_2$ (see Table S1). Using the strategy proposed by Fu \emph{et al}. \cite{R42}, the topological invariant of siliconeet is determined to be $\emph{Z}_2$ = 1, which can be ascribed to the band inversion near the Y point [Fig. \ref{fig4}(a)]. Moreover, the SOC gap of siliconeet, 2.58 meV [Fig. \ref{fig4}(b)], is much larger than that of silicene, due to the lower symmetry of its unique buckled configuration with fivefold and fourfold coordinations. The energetic preferability and appreciable nontrivial band gap superior to silicene make siliconeet a promising candidate for realizing the QSH effect.

In conclusion, based on systematic evolutionary structure searching, we predict the stable 2D-Si Dirac allotrope named siliconeet. Its peculiar pentagonal rings and unique fivefold coordination are demonstrated to play a critical role in the novel electronic properties. Its direction-dependent Dirac cones are further proved to be robust against strain. Siliconeet is also confirmed to be a 2D topological insulator with an appreciable bandgap induced by spin-orbit coupling. In future, it may become a potential candidate to be integrated in Si-based electronic technology. Our findings not only extend the family of Dirac semimetals, but also clarify the structural stability for 2D forms of silicon. \\

Z. W. thanks the China Scholarship Council (No. 201408320093), the Natural Science Foundation of Jiangsu Province (Grant No. BK20130859), the University Natural Science Research Project of Jiangsu Province (Grant No. 13KJB510019) and NUPTSF (Grants No. NY213010 and NY214030); M. Z. thanks the support from the National Basic Research Program of China (Grant No. 2012CB932302) and the National Natural Science Foundation of China (Grant No. 91221101); X. F. Z. thanks the National Science Foundation of China (Grant No. 11174152), the National 973 Program of China (Grant No. 2012CB921900), the Program for New Century Excellent Talents in University (Grant No. NCET-12-0278), and the Fundamental Research Funds for the Central Universities (Grant No. 65121009); A. R. O. thanks the Government of Russian Federation (No. 14.A12.31.0003), the Foreign Talents Introduction and Academic Exchange Program (No. B08040) and the support from SUNY 4E NoE.




\begin{references}

\bibitem{R01} K. S. Novoselov, A. K. Geim, S. V. Morozov, D. Jiang, Y. Zhang, S. V. Dubonos, I. V. Grigorieva, and A. A. Firsov, Science \textbf{306}, 666 (2004).

\bibitem{R02} A. H. Castro Neto, F. Guinea, N. M. R. Peres, K. S. Novoselov, and A. K. Geim, Rev. Mod. Phys. \textbf{81}, 109 (2009).

\bibitem{R03} K. I. Bolotin, K. J. Sikes, Z. Jiang, M. Klima, G. Fudenberg, J. Hone, P. Kim, and H. L. Stormer, Solid State Comm. \textbf{146}, 351 (2008).

\bibitem{R04} C. L. Kane and E. J. Mele, Phys. Rev. Lett. \textbf{95}, 226801 (2005).

\bibitem{R05} Y. Zhang, Y.-W. Tan, H. L. Stormer, and P. Kim, Nature \textbf{438}, 201 (2005).

\bibitem{R06} K. I. Bolotin, F. Ghahari, M. D. Shulman, H. L. Stormer, and P. Kim, Nature \textbf{462}, 196 (2009).

\bibitem{R07} K. Takeda and K. Shiraishi, Phys. Rev. B \textbf{50}, 14916 (1994).

\bibitem{R08} G. G. Guzm\'{a}n-Verri and L. C. Lew Yan Voon, Phys. Rev. B \textbf{76}, 075131 (2007).

\bibitem{R09} S. Cahangirov, M. Topsakal, E. Akt\"{u}rk, H. \c{S}ahin, and S. Ciraci, Phys. Rev. Lett. \textbf{102}, 236804 (2009).

\bibitem{R10} C.-C. Liu, W. Feng, and Y. Yao, Phys. Rev. Lett. \textbf{107}, 076802 (2011).

\bibitem{R11} F. Liu, C.-C. Liu, K. Wu, F. Yang, and Y. Yao, Phys. Rev. Lett. \textbf{111}, 066804 (2013).

\bibitem{R12} C. Xu, G. Luo, Q. Liu, J. Zheng, Z. Zhang, S. Nagase, Z. Gao, and J. Lu, Nanoscale \textbf{4}, 3111 (2012).

\bibitem{R13} M. Ezawa, Phys. Rev. Lett. \textbf{109}, 055502 (2012).

\bibitem{R14} E. Cinquanta, E. Scalise, D. Chiappe, C. Grazianetti, B. van den Broek, M. Houssa, M. Fanciulli, and A. Molle, J. Phys. Chem. C \textbf{117}, 16719 (2013).

\bibitem{R15} P. Vogt, P. De Padova, C. Quaresima, J. Avila, E. Frantzeskakis, M. C. Asensio, A. Resta, B. Ealet, and G. Le Lay, Phys. Rev. Lett. \textbf{108}, 155501 (2012).

\bibitem{R16} B. Feng, H. Li, C.-C. Liu, T.-N. Shao, P. Cheng, Y. Yao, S. Meng, L. Chen, and K. Wu, ACS Nano \textbf{7}, 9049 (2013).

\bibitem{R17} S. K. Mahatha, P. Moras, V. Bellini, P. M. Sheverdyaeva, C. Struzzi, L. Petaccia, and C. Carbone, Phys. Rev. B \textbf{89}, 201416 (2014).

\bibitem{R18} B. Aufray, A. Kara, S. Vizzini, H. Oughaddou, C. L\'{e}andri, B. Ealet, and G. Le Lay, Appl. Phys. Lett. \textbf{96}, 183102 (2010).

\bibitem{R19} P. De Padova, C. Quaresima, C. Ottaviani, P. M. Sheverdyaeva, P. Moras, C. Carbone, D. Topwal, B. Olivieri, A. Kara, H. Oughaddou, B. Aufray, and G. Le Lay, Appl. Phys. Lett. \textbf{96}, 261905 (2010).

\bibitem{R20} L. Meng, Y. Wang, L. Zhang, S. Du, R. Wu, L. Li, Y. Zhang, G. Li, H. Zhou, W. A. Hofer, and H.-J. Gao, Nano Lett. \textbf{13}, 685 (2013).

\bibitem{R21} A. Fleurence, R. Friedlein, T. Ozaki, H. Kawai, Y. Wang, and Y. Yamada-Takamura, Phys. Rev. Lett. \textbf{108}, 245501 (2012).

\bibitem{R22} D. Chiappe, E. Scalise, E. Cinquanta, C. Grazianetti, B. van den Broek, M. Fanciulli, M. Houssa, and A. Molle, Adv. Mater. \textbf{26}, 2096 (2014).

\bibitem{R23} C.-L. Lin, R. Arafune, K. Kawahara, M. Kanno, N. Tsukahara, E. Minamitani, Y. Kim, M. Kawai, and N. Takagi, Phys. Rev. Lett. \textbf{110}, 076801 (2013).

\bibitem{R24} Z.-X. Guo, S. Furuya, J.-i. Iwata, and A. Oshiyama, Phys. Rev. B \textbf{87}, 235435 (2013).

\bibitem{R25} M. X. Chen and M. Weinert, Nano Lett. \textbf{14}, 5189 (2014).

\bibitem{R26} A. Molle, C. Grazianetti, D. Chiappe, E. Cinquanta, E. Cianci, G. Tallarida, and M. Fanciulli, Adv. Func. Mater. \textbf{23}, 4340 (2013).

\bibitem{R27} D. Malko, C. Neiss, F. Vi\~{n}es, and A. G\"{o}rling, Phys. Rev. Lett. \textbf{108}, 086804 (2012).

\bibitem{R28} X.-F. Zhou, X. Dong, A. R. Oganov, Q. Zhu, Y. Tian, and H.-T. Wang, Phys. Rev. Lett. \textbf{112}, 085502 (2014).

\bibitem{R29} Z. Wang, X.-F. Zhou, X. Zhang, Q. Zhu, H. Dong, M. Zhao, and A. R. Oganov, Nano Lett. \textbf{15}, 6182 (2015).

\bibitem{R30} A. R. Oganov and C. W. Glass, J. Chem. Phys. \textbf{124}, 244704 (2006).

\bibitem{R31} C. W. Glass, A. R. Oganov, and N. Hansen, Comput. Phys. Commun. \textbf{175}, 713 (2006).

\bibitem{R32} Q. Zhu, L. Li, A. R. Oganov, and P. B. Allen, Phys. Rev. B \textbf{87}, 195317 (2013).

\bibitem{R33} P. E. Bl\"{o}chl, Phys. Rev. B \textbf{50}, 17953 (1994).

\bibitem{R34} G. Kresse and J. Furthmuller, Phys. Rev. B \textbf{54}, 11169 (1996).

\bibitem{R35} G. Kresse and J. Furthmuller, Comput. Mater. Sci. \textbf{6}, 15 (1996).

\bibitem{R36} J. P. Perdew, K. Burke, and M. Ernzerhof, Phys. Rev. Lett. \textbf{77}, 3865 (1996).

\bibitem{R37} J. Heyd, G. E. Scuseria, and M. Ernzerhof, J. Chem. Phys. \textbf{118}, 8207 (2003).

\bibitem{R38} J. Heyd, G. E. Scuseria, and M. Ernzerhof, J. Chem. Phys. \textbf{124}, 219906 (2006).

\bibitem{R39} A. Togo, F. Oba, and I. Tanaka, Phys. Rev. B \textbf{78}, 134106 (2008).

\bibitem{R40} S. Botti, J. A. Flores-Livas, M. Amsler, S. Goedecker, and M. A. L. Marques, Phys. Rev. B \textbf{86}, 121204 (2012).

\bibitem{R41} Q. Wang, B. Xu, J. Sun, H. Liu, Z. Zhao, D. Yu, C. Fan, and J. He, J. Am. Chem. Soc. \textbf{136}, 9826 (2014).

\bibitem{R42} L. Fu, and C. L. Kane, Phys. Rev. B \textbf{76}, 045302 (2007).

\end{references}
\end{document}